\documentclass[conference]{IEEEtran}
\ifCLASSINFOpdf
\else
\fi

\usepackage{cite}
\usepackage{amsmath,amssymb,amsfonts}
\usepackage{caption}
\usepackage{subcaption}
\usepackage{graphicx}
\usepackage{textcomp}
\usepackage{xcolor}
\usepackage{amsmath}
\usepackage{multirow}
\usepackage{booktabs}
\usepackage{amsthm}
\usepackage{stfloats}
\usepackage{booktabs}
\usepackage{algorithm}
\usepackage{algorithmic}
\usepackage{amsmath}
\usepackage{url}

\begin{document}
%
\title{Achieving Deterministic Service in Mobile Edge Computing (MEC) Networks}

 \author{\IEEEauthorblockN{Binwei Wu\IEEEauthorrefmark{1},
 Jiasen Wang\IEEEauthorrefmark{1},
 Yanyan Wang\IEEEauthorrefmark{1},
 Weiqian Tan\IEEEauthorrefmark{3}\IEEEauthorrefmark{1},
 Yudong Huang\IEEEauthorrefmark{2}\IEEEauthorrefmark{1}}

 \IEEEauthorblockA{\IEEEauthorrefmark{1} Purple Mountain Laboratories, Nanjing, China\\}

 \IEEEauthorblockA{\IEEEauthorrefmark{2} Beijing University of Posts and Telecommunications, Beijing, China\\}

 \IEEEauthorblockA{\IEEEauthorrefmark{3} Southeast University, Nanjing, China\\}}

\maketitle

\begin{abstract}
Mobile edge computing (MEC) is proposed to boost high-efficient and time-sensitive 5G applications. However, the ``microburst'' may occur even in lightly-loaded scenarios, which leads to the indeterministic service latency (i.e., unpredictable delay or delay variation), hence hindering the deployment of MEC. Deterministic IP networking (DIP) has been proposed that can provide bounds on latency, and high reliability in the large-scale networks. Nevertheless, the direct migration of DIP into the MEC network is non-trivial owing to its original design for the Ethernet with homogeneous devices. Meanwhile, DIP also faces the challenges on the network throughput and scheduling flexibility. In this paper, we delve into the adoption of DIP for the MEC networks and some of the relevant aspects. A deterministic MEC (D-MEC) network is proposed to deliver the deterministic service (i.e., providing the MEC service with bounded service latency). In the D-MEC network, two mechanisms, including the cycle mapping and cycle shifting, are designed to enable: (i) seamless and deterministic transmission with heterogeneous underlaid resources; and (ii) traffic shaping on the edges to improve the resource utilization. We also formulate a joint configuration to maximize the network throughput with deterministic QoS guarantees. Extensive simulations verify that the proposed D-MEC network can achieve a deterministic MEC service, even in the highly-loaded scenarios.
\end{abstract}


%
\IEEEpeerreviewmaketitle

\section{Introduction}
Mobile edge computing (MEC) is proposed to empower delay-sensitive and resource-hungry applications, such as mobile gaming, automatic remote control. These applications can offload their tasks that have strict requirements on the service latency to the nearby MEC servers~\cite{mao2017survey}.
Conventional works may tempt to think that short communication distances plus lightly-loaded communication links in the MEC networks will yield low service latency and small end-to-end jitter. This is unfortunately not the case. The trivial experiments have shown that significant latency and jitter could be experienced even in a very lightly-loaded network due to the ``microburst''~\cite{charny2000delay}. 
The unpredictable and indeterministic service latency in the MEC networks is seen as a major impediment to enable time-sensitive applications.

Internet Engineering Task Force (IETF) has developed a collection of standards, known as Deterministic Networking (DetNet), that can provide bounds on latency, packet delay variation (jitter), and high reliability in the large-scale networks~\cite{nasrallah2018ultra}. Candidate data-plane mechanisms, such as deterministic IP networking (DIP), cycle specified queuing and forwarding (CSQF), have been proposed. In these mechanisms, the nodes (network devices) have synchronized frequency, and each node forwards packets in a slotted fashion based on cycle identifiers carried in packets. It has been shown that DIP can provide a delay-bounded transmission service (maximum jitter $\leq 2 \Delta_{\rm{dip}}$) in large-scale networks with no packet loss, where $\Delta_{\rm{dip}}$ is the cycle length in the DIP-enabled routers~\cite{QiangSDF}.

It is worthy to integrate the DIP into the MEC networks so that the networks can provide the deterministic service latency. However, the migration of DIP is non-trivial owing to its original design for the ethernet networks with homogeneous devices. The mechanisms in DIP, such as equal time division, are no longer optional in the MEC networks~\cite{LiQIANG:12}. Thus, further extensions need to be made for the compatibility with heterogeneous devices, e.g., mobile devices, APs, routers, and MEC servers. Meanwhile, the performance of DIP under complicated topologies requires further investigation. The cycle-based forwarding mechanism in the DIP is flow-independent, which leads to low resource utilization in most cases~\cite{9155434}.

To solve the above-mentioned issues, this paper delves into the adoption of DIP for the MEC networks and some of the relevant aspects. We exploit the DIP technique and propose a deterministic MEC (D-MEC) network. The D-MEC network can bound the service latency from an end-to-end perspective, i.e., the service latency refers to the time taken for a task from its generation until its end of processing. For the compatibility of heterogeneous resources (e.g., wireless radio blocks, link bandwidth, CPU cycles), the network is logically divided into multiple domains with different configurations. We redefine the cycle mapping mechanism as a function to achieve seamless transmission across multiple domains. By proposing a cycle shifting mechanism, we shape the traffic at the edge of each domain, which enhances the scheduling flexibility and the network throughput. Then, a joint configuration approach is proposed that enables the optimization from an end-to-end perspective. Extensive simulations show that the proposed D-MEC can achieve a deterministic service. Meanwhile, through the end-to-end optimization, the network throughput (i.e., the number of acceptable demands) is improved.

\section{Background and Related works}
\subsection{Deterministic IP networking}
DetNet is formed to address the deterministic transmission in the large-scale networks. Among all the candidate mechanisms in DetNet, DIP appears as one of the most top choices owing to its simplicity and efficiency.

\begin{figure}[ht]
\centering
\includegraphics[width=.80\linewidth]{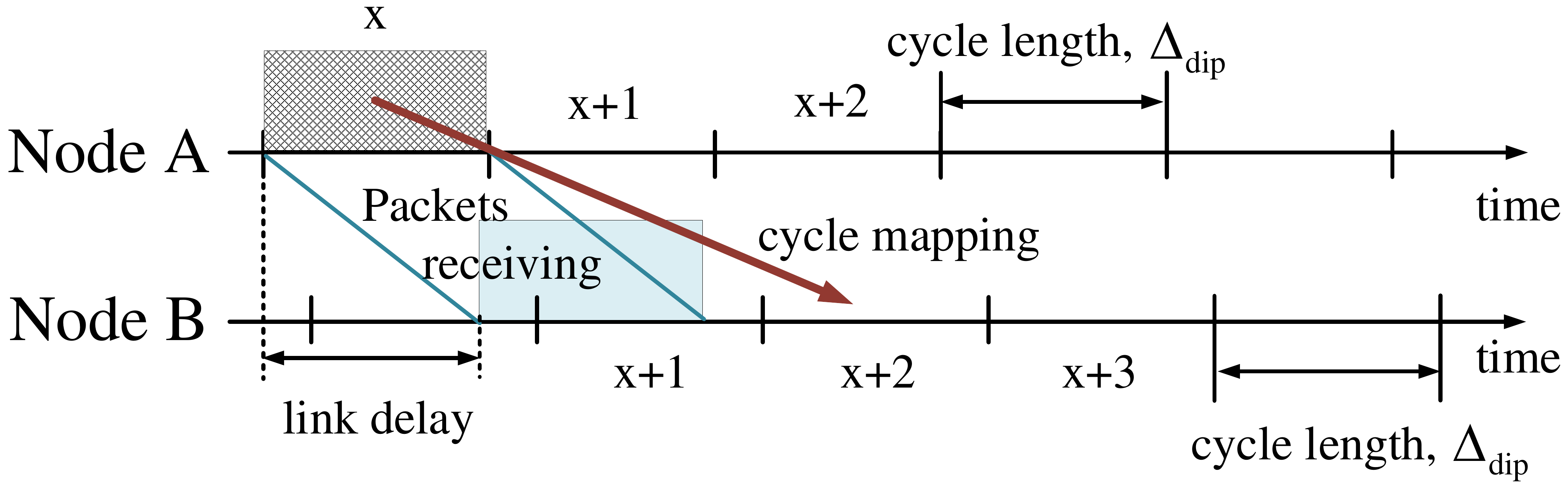}
\caption{\small Frequency synchronization and cycle mapping in the DIP.}
\label{fig:dip}
\end{figure}

In the DIP-enabled networks, all the nodes achieve {\bf{frequency synchronization}}: nodes maintain the same clock frequency $1/\Delta_{\rm{dip}}$, as shown in Fig.\ref{fig:dip}. Due to the offset of the cycles' start time and long link delay, the downstream Node B may receive the packets from Node A at two different cycles (cycle $x$ and $x+1$). To absorb this variation, DIP defines the {\bf{cycle mapping mechanism}}. Essentially, it involves a constant offset of the cycles, which specifies that at which cycle the downstream node should re-send the packets out. For example, a cycle mapping $x \to x+2$ requires that the packets from Node A at cycle $x$ should be re-sent out by Node B at cycle $x+2$. Consequentially, the packets at cycle $x+1$ on Node A are required to be sent out by Node B at cycle $x+3$ (owing to the identical cycle length $\Delta_{\rm{dip}}$).
An explicit cycle mapping needs to be maintained for any pair of neighboring nodes. Through the hop-by-hop cycle mapping-based forwarding, the precise knowledge of the position of a packet inside the network at a generic time $t$ is known, hence ensuring the deterministic transmission.
The detail of DIP can be referred to~\cite{QiangSDF} and~\cite{LiQIANG:12}.

\subsection{Related works}
Extensional works have been done on exploiting the deterministic techniques to enhance the reliability and throughput of the networks.
In standardization, working groups are formed, including, TSN, DetNet, and 6TiSCH~\cite{nasrallah2018ultra}. Both TSN and DetNet contain mechanisms, allowing for the coexistence of different traffic classes with different priorities on the wired networks. 6TiSCH focuses on providing strict reliability in wireless IoT scenarios. The solutions are based on Time-Synchronized Channel Hopping, a medium access control technique at the heart of industrial standards~\cite{vilajosana2019ietf}.

In the research field, the authors of \cite{8736787, 7858136} adopt the time-aware shaper (TAS, a candidate shaping mechanism in TSN) on the fronthual links to guarantee the delay requirement of high-priority traffic flows. Similarly, the authors of \cite{9373015} incorporate the TSN into the 5G backhual links which enable time-critical services and their coexistence with other conventional network flows. A TSN-based control plane is constructed in \cite{9410215}, which interconnects the slice control and management system of the mobile network. Nevertheless, TSN is essentially a Layer 2 technique and unsuitable for the scenarios with large geographic coverage~\cite{nasrallah2018ultra}. Thus, we exploit the DIP techniques in MEC networks. As an extension of cyclic queuing and forwarding, DIP exhibits the problem of low resource utilization~\cite{9155434}. The authors of \cite{9473691} and \cite{krolikowski2021joint} adopt the load balance and traffic shaping mechanism to improve the throughput. However, these mechanisms cannot be directly adopted in MEC owing to its original design with homogeneous devices/resources.

\subsection{MEC networks}
\begin{figure}[ht]
\centering
\includegraphics[width=.80\linewidth]{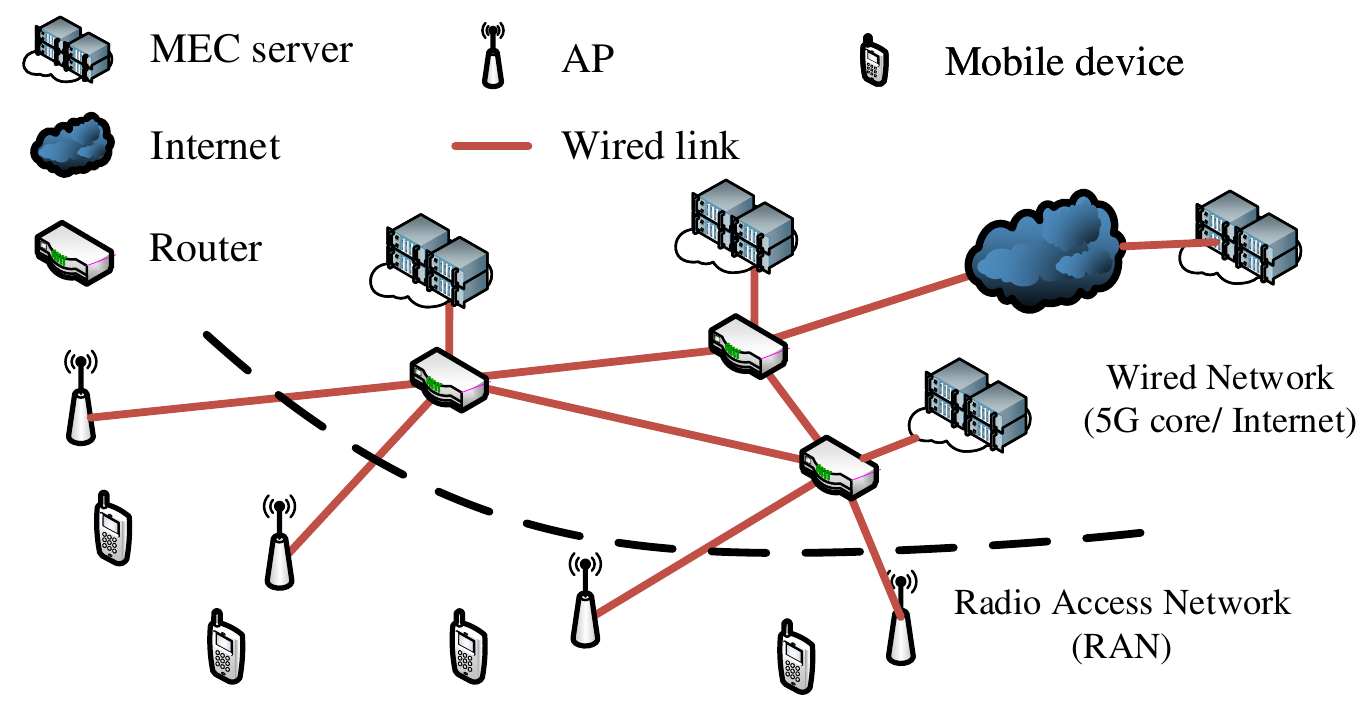}
\caption{\small A typical MEC network comprises a RAN and a wired network. Some devices equipped with computational resources (referred to MEC servers) can provide MEC services.}
\label{fig:NetworkModel}
\end{figure}

We use $\mathcal{G} = \left\{\mathcal{V}, \mathcal{E}\right\}$ to denote the MEC network, where $\mathcal{V}$ and $\mathcal{E}$ are set of nodes and links, respectively. The nodes include mobile devices ($\mathcal{U}$), APs ($\mathcal{V}_{\rm{ap}}$), routers ($\mathcal{V}_{\rm{rou}}$), and MEC servers ($\mathcal{V}_{\rm{mec}}$). Each MEC server $j \in \mathcal{V}_{\rm{mec}}$ is equipped with limited computational resources. The capacity of server $j$ is ${\rm{BW}}_j^{\rm{mec}}$, measured by CPU cycles per second. 

A schedule-path (s-path) ${p} = \left(v_0, v_1, \cdots, v_{|p|}\right) \in \mathcal{P}$ is a path from the mobile devices to the MEC servers. $p$ contains a set of nodes, where $v_0 \in \mathcal{U}$, $v_1 \in \mathcal{V}_{\rm{ap}}$, and $v_{|p|} \in \mathcal{V}_{\rm{mec}}$. Initially, the tasks of users (i.e., MEC applications) are generated on the mobile devices and aggregated on their associated APs through wireless links. We simplify the complicated 3GPP processes and quantize the wireless transmission resources in a time-frequency grid, named as resource blocks (RBs). Each RB can only be assigned to a single user at a time. The minimum reserved capacity of a RB is ${\rm{BW}}^{\rm{res}}_{c, f}$, where $c$ and $f$ are the index of time and frequency band, respectively. Then, the APs forward the tasks to MEC servers through routers and wired links. A wired link $e = (v_i, v_j)\in\mathcal{E}_{\rm{wired}}$ incurs a link delay $\tau_{(v_i, v_j)}$ and its bandwidth is ${\rm{BW}}_{(v_i, v_j)}^{\rm{link}}$, measured by bits per second.

This paper only considers the deterministic periodic applications, i.e., the tasks are generated with known arrival time, packet size, and periodicity. To describe the pattern of the tasks, the time on the mobile devices is divided into the time cycles with duration $\Delta_{\rm{tti}}$ (named as the transmission time interval cycles or TTI cycles). Then, we describe the task flow of a user as a demand $d\in\mathcal{D}$, defined by a 5-tuple $\left<s^d, T^d, c^d, \omega^d, \Gamma^d \right>$, where $s^d\in\mathcal{U}$ is its source, $T^d$ is its cyclic time, $c^d$ indicates its arrival time (i.e., the index of TTI cycles), $\omega^d$ (in bits) is the payload of  an individual task, and $\Gamma^d$ is the maximum tolerable service latency. Each demand $d$ is destined for a MEC server that needs to be properly decided. Meanwhile, the tasks consume $\kappa$ CPU cycles per bit on the MEC server.

\section{Network design}

\subsection{The overview of D-MEC networks}

The procedure of acquiring the delay-bounded MEC service is listed as follows.

1. We logically divide the network into domains (i.e., RAN-D, WN-D, and MECS-D), where the nodes partition the time into cycles and carry out the synchronization, accordingly. All the nodes in the D-MEC network must support the deterministic mechanisms, e.g., cycle mapping and cyclic shifting.

2. Each port in the network installs a series of cyclic queues, i.e., $Q$ queues and $Q \geq 3$. Every cyclic queue corresponds to a time cycle. Initially, a cycle mapping relationship is established for each pair of neighboring nodes.

3. Admission control is conducted for every arriving demand. For each accepted demand $d$, the controller decides its s-path $p^d$, associated MEC server, and the resources (i.e., RBs, link bandwidth, and computational resources).

4. Every demand $d$ must yield to its pre-defined deterministic pattern, i.e., $\left<s^d, T^d, c^d, \omega^d, \Gamma^d \right>$. If 1, 2, 3 are satisfied, the network provides the D-MEC service, as follows: 

\begin{itemize}
    \item[*] Dedicated RBs are allocated in advance. Packets would be buffered on the mobile devices until their transmitted time intervals arrive.
    \item[*] By assigning different delays (indicated by the cycle shifting), the APs reshape the traffic on a per-demand granularity, which balances the workloads of cycles on the downstream links.
    \item[*] The routers forward the tasks with the DIP mechanism, which provides a delay-bounded transmission on the wired links (i.e., from the APs to the MEC servers).
    \item[*] The MEC servers also shape the tasks with the cycle shifting mechanism. An additional delay is assigned to balance the workloads of different intervals on the CPUs.
\end{itemize}

5. D-MEC leverages the SR label stack to implement the cycle mapping and cycle shifting. In the SR label stack, each packet carries a cycle identifier of the sending node. The downstream node will check the cycle mapping relationship table and the cycle shifting information, swap the cycle identifier with a new cycle identifier, then put the packet into the appropriate queue, waiting for the future transmission.

The key techniques are elaborated as follows.

\subsection{Network domains and time partition}\label{section3.2}

The D-MEC network is logically divided into RAN-D, WN-D, and MECS-D, as follows.

{\textit{RAN-D}:} It contains mobile devices and APs. The time is divided into TTI cycles with length $\Delta_{\rm{tti}}$. In accord with the 5G standards, typical values of $\Delta_{\rm{tti}}$ are 0.5~ms, 0.125~ms, etc.

{\textit{WN-D}:} It consists of  the devices (e.g., routers) that provide wired connectivity. We partition the time into DIP cycles with length $\Delta_{\rm{dip}}$. A typical value for $\Delta_{\rm{dip}}$ is 20 $\mu$s.

{\textit{MECS-D}:} It contains all the MEC servers. The time is divided into computation cycles with length $\Delta_{\rm{mec}}$. Let $\Delta_{\rm{mec}} \geq \max\nolimits_{d} \{\kappa^d\omega^d / {\rm{BW}}^{\rm{mec}}_j\}, \forall j\in\mathcal{V}_{\rm{mec}}$ so that any demand $d\in\mathcal{D}$ can be processed in a single computation cycle.

{\textit{Edge of the domains: }}
The APs and MEC servers are located at the edge of RAN-D and MECS-D, respectively.


{\textit{Definition of hypercycle: }} We define the ``hypercycle'', given by
\begin{equation}\label{e8}
\Delta_{\rm{hc}} =  N_{\rm{hc}} \; {\rm{lcm}} \left(\{\Delta_{\rm{tti}}, \Delta_{\rm{dip}}, \Delta_{\rm{mec}}\} \cup \{T^{d}, \forall d\in\mathcal{D}\}\right)
\end{equation}
where ${\rm{lcm}} (\cdot)$ represents the least common multiple, and $N_{\rm{hc}}\in\mathbb{Z}_{+}$. We use $\mathcal{N}_{\rm{tti}}$, $\mathcal{N}_{\rm{dip}}$, and $\mathcal{N}_{\rm{mec}}$ to denote the set of TTI cycles, DIP cycles, and computation cycles in a single hypercycle, respectively. In this paper, the cycle index starts on 0. Let $N_{\rm{tti}} = |\mathcal{N}_{\rm{tti}}|, N_{\rm{dip}} = |\mathcal{N}_{\rm{dip}}|, N_{\rm{mec}} = |\mathcal{N}_{\rm{mec}}|$. Then,  $\Delta_{\rm{tti}}N_{\rm{tti}} = \Delta_{\rm{dip}}N_{\rm{dip}} = \Delta_{\rm{mec}}N_{\rm{mec}} = \Delta_{\rm{hc}}$. $N_{\rm{hc}}$ is sufficiently large so that $N_{\rm{dip}} \geq Q$ and $N_{\rm{mec}} \geq Q$. For the simplicity, we let $\{T^{d} = \Delta_{\rm{hc}}, \forall d\in\mathcal{D}\}$.

Since the traffic pattern of time-sensitive flows repeats for every realization of $\Delta_{\rm{hc}}$, the hypercycle is the resource allocation window that we are considering. 


\begin{figure}[ht]
\centering
\includegraphics[width=.90\linewidth]{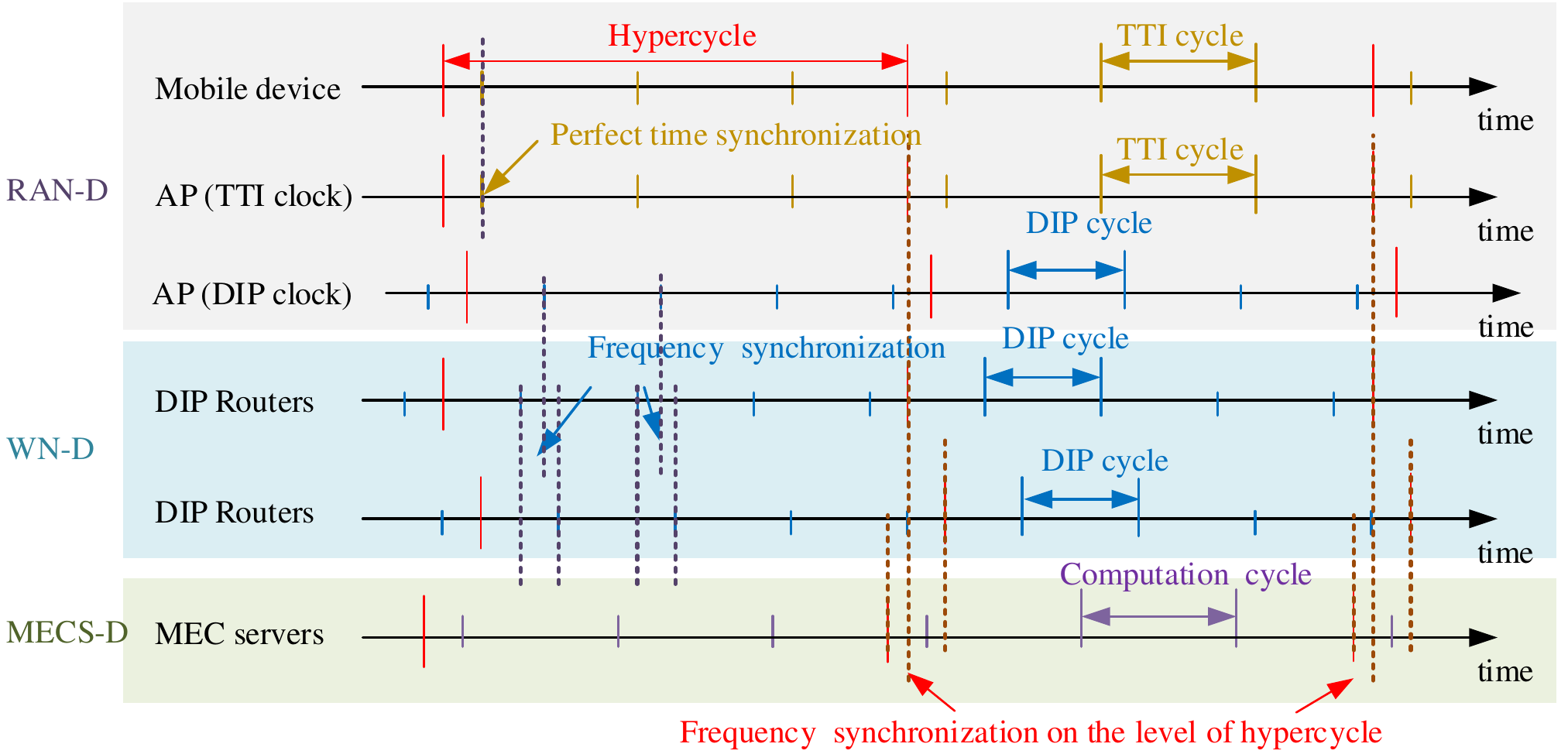}
\caption{\small Time partition and synchronization in different domains.  }
\label{fig:time-domains}
\end{figure}

%

\subsection{Synchronization}\label{section3.3}

The nodes in the D-MEC network synchronize with each other to establish a unified understanding of time. Given the features of different domains, we use different synchronization mechanisms.

{\textit{RAN-D:}} Each node installs a TTI clock with length $\Delta_{\rm{tti}}$. The APs achieve strict time synchronization with the mobile devices in their coverage, as required in the 5G standards.

{\textit{WN-D}:} We only require the frequency synchronization due to the large geographical coverage: every node (e.g., router) maintains the same clock frequency $1/\Delta_{\rm{dip}}$ (i.e., the DIP clock), but do not require the same start time.

{\textit{MECS-D}:} Every MEC server maintains a local clock with duration $\Delta_{\rm{mec}}$. No time synchronization is considered among the MEC servers.

{\textit{Edge of the domains:}}
The APs also install DIP clocks, which achieve the frequency synchronization with their neighboring routers. The MEC servers achieve the frequency synchronization with their neighboring routers on the level of hypercycle, i.e., the offsets of the start time of the hypercycle (denoted as $\tau_{v_i,v_j}^{\rm{hc}}$, where $v_i\in\mathcal{V}_{\rm{rou}}, v_j\in\mathcal{V}_{\rm{mec}}$) are constant. Let $\tau_{v_i,v_j}^{\rm{hc}} = t_{v_j}^{\rm{hc}} - t_{v_i}^{\rm{hc}}$, where $t_{v_i}^{\rm{hc}}$ and $t_{v_j}^{\rm{hc}}$ is the start time of hypercycle on node $v_i$ and $v_j$, respectively.

\subsection{Cycle mapping in WN-D and MECS-D}\label{section3.4}

Instead of specifying when (or which cycle) to re-send the tasks out (as defined in the DIP), the proposed cycle mapping outputs the last possible receiving cycle on the downstream node.

\begin{figure}[ht]
\centering
\includegraphics[width=.90\linewidth]{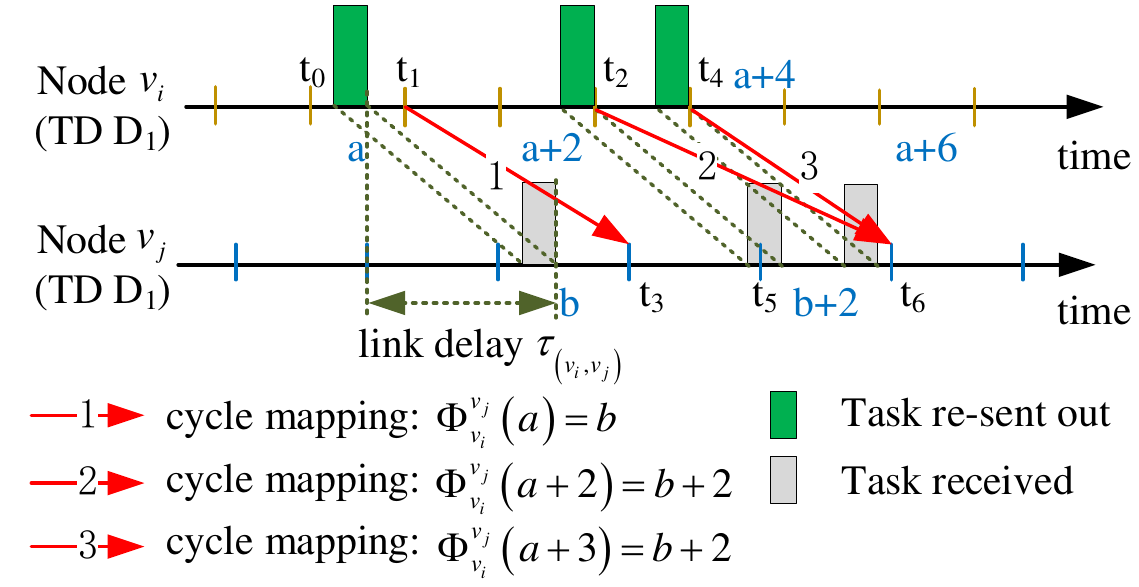}
\caption{\small Cycle mapping from node $v_i$ (in domain $D_1$) to node $v_j$ (in domain $D_2$). $t_0-t_6$ is the absolute time in the real world. Cycle mapping $\Phi_{v_i}^{v_j}(a)=b$ represents that the tasks that sent out at cycle $a$ (from $t_0$ to $t_1$) on node $v_i$ would definitely be received by the end of cycle $b$ (i.e., $t_3$) on node $v_j$ . }
\label{fig:cyclemapping}
\end{figure}

Consider the cycle mapping on two neighboring nodes (e.g., $v_i$ and $v_j$), where $v_i$ and $v_j$ locate in the domains $D_i$ and $D_j$, respectively. If a task is sent out at cycle $a\in\mathcal{N}_{D_i}$ on node $v_i$, the index of the last possible receiving cycle on node $v_j$ depends on: (i) the link delay $\tau_{(v_i, v_j)}$; (ii) the length of hypercycle $\Delta_{\rm{hc}}$; (iii) the cycle length  $\Delta_{D_i}$ and $\Delta_{D_j}$; and (iv) the offset of hypercycles on node $v_i$ and $v_j$, $\tau_{v_i, v_j}^{\rm{hc}}$. The proposed cycle mapping from node $v_i$ to $v_j$ is a function, i.e., $\Phi_{v_i}^{v_j}:\mathcal{N}_{D_i}\to \mathcal{N}_{D_j}$, given by
\begin{equation}
\Phi_{v_i}^{v_j}(a) = {\rm{mod}}\left(\lfloor \frac {(a+1)\Delta_{D_i}+\tau_{(v_i,v_j)}-\tau^{\rm{hc}}_{v_i,v_j}}{\Delta_{D_j}}\rfloor, N_{D_{j}}\right)
\end{equation}
where ${\rm{mod}}(\cdot)$ is the modulo operation.
Mapping $\Phi_{v_i}^{v_j}(\cdot)$ is a periodic function, i.e., $
\Phi_{v_i}^{v_j}(a+k N_{D_i}) = \Phi_{v_i}^{v_j}(a), \forall a\in\mathcal{N}_{D_i}$,
where $ k\in\mathbb{Z}_{+}$ and $ N_{D_i}=|\mathcal{N}_{D_i}|$ since node $v_i$ and $v_j$ can always achieve the frequency synchronization on the level of hypercycle.

We define a function $\phi_{v_i}^{v_j}: \mathcal{N}_{D_i}\to\mathbb{R}_{\geq 0}$, give by
\begin{equation}\label{phi}
\phi_{v_i}^{v_j}(a) = \lfloor \frac {(a+1)\Delta_{D_i}+\tau_{(v_i, v_j)}-\tau^{\rm{hc}}_{v_i,v_j}}{\Delta_{D_j}}\rfloor
 \Delta_{D_j} + \tau^{\rm{hc}}_{v_i, v_j} - a \Delta_{D_j}.
\end{equation}
$\phi_{v_i}^{v_j}(a)$ outputs the delay due to mapping. Then, the worst-case transmission delay for the packets sent out at cycle $a$ on node $v_i$ can be derived by $\phi_{v_i}^{v_j}(a) + \Delta_{D_i}$.

\subsection{Cycle shifting in WN-D and MECS-D}

We use the cycle shifting to specify when (or which cycle) to re-send the tasks out. A cycle shift is an explicit additional delay (expressed in cycles)  introduced at the intermediate nodes. We carry out the cycle shifting on a per-demand granularity.

\begin{figure}[ht]
\centering
\includegraphics[width=.90\linewidth]{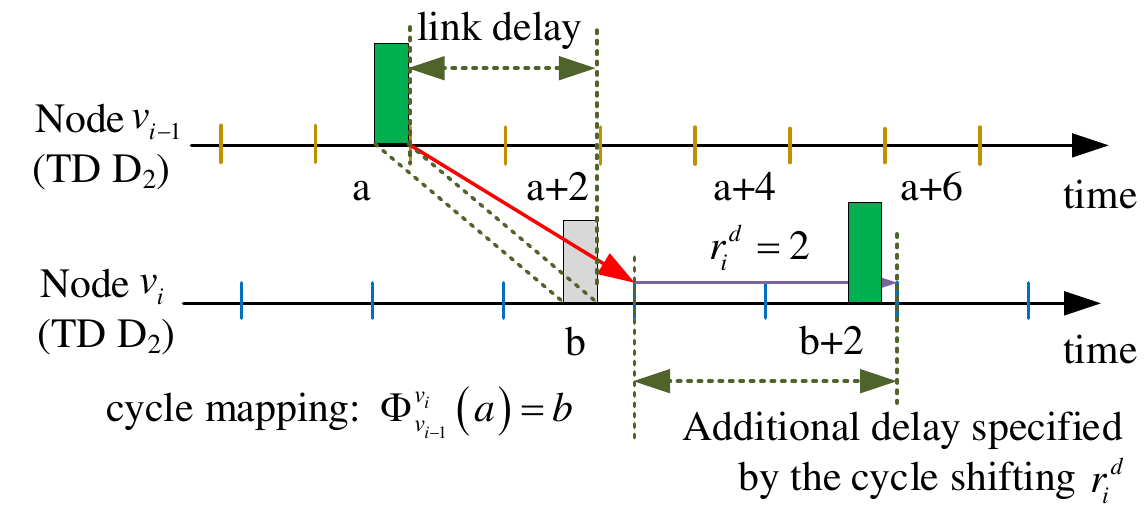}
\caption{\small Cycle shifting mechanism in WN-D and MECS-D. Shift $r^d_i$ specify an additional delay on $i^{\rm{th}}$ intermediated node of s-path $p^d$.}
\label{fig:shifting}
\end{figure}

Let $p^d = (v_0, \cdots, v_{|p^d|})$ be the s-path assigned to demand $d$. A segment of $p^d$ in the WN-D and MECS-D is $\bar p^d = (v_2, \cdots, v_{|p^d|})$. We associate a cycle shifting vector $\bar{\bf{r}}^d_p = (r^d_2, \cdots, r^d_{|p^d|})$, where $r_i^d \in \mathbb{Z}_{\geq 1}, \forall i \in \{2,\cdots,|p^d|\}$.
As Fig.~\ref{fig:shifting} illustrated, if the tasks of demand $d$ would arrive the node $v_i$ by the end of cycle $b$ in every realization of the hypercycle (indicated by the cycle mapping), then a cycle shift of $r_i^d$ means that the tasks would be re-send out at cycle $\left(b + r_i^d\right)$ on node $v_i$.

To implement the cycle shifting mechanism, each port maintains more queues, i.e., $Q \geq 3$ (1 sending queue and $Q-1$ receiving queues).
A shift $r_i^d \geq 1$ is achieved by scheduling the tasks into a specific receiving queue. The maximum number of shifts at a node is $R^d_i = Q-2$.

{\textit{Cycle shifting inside the WN-D:}} WN-D use the DIP mechanism to forward the tasks: if the last possible receiving cycle of the task on node $v_i$ is cycle $a$; then, the task would be re-sent out at cycle $a+1$. Thus, we have $r_i^d = 1, \forall i\in \{2, \cdots, |p^d|-1\}, \forall d\in\mathcal{D}$.

{\textit{Cycle shifting inside the MECS-D:}} The computational cycles on MEC servers represent different time intervals for the task processing. If the last possible receiving (computational) cycle of the task on server $v_{|p^d|}\in\mathcal{V}_{\rm{mec}}$ is the cycle $a$. Then, a shift $r^d_{|p^d|}$ means that the task would be processed during the cycle $a+r^d_{|p^d|}$.

\subsection{Cycle mapping and shifting in the RAN-D}

The deterministic transmission in the RAN-D is guaranteed by assigning the dedicated RBs.
Let ${\mathcal{Y}} = \{{\bf{y}}^d\}_{d\in\mathcal{D}}$ present the RB assignment of all the demands, where ${\bf y}^d = ({y}^d_{c,f})_{c\in\mathcal{N}_{\rm{tti}}, f\in\mathcal{F}}$. If RB $(c,f)$ is assigned to demand $d$, ${y}^d_{c,f}=1$; otherwise, ${y}^d_{c,f}=0$.
For every accepted demand $d$, we must guarantee that all of its assigned RBs are available for every realization of the hypercycle.
The tasks would be buffered on the mobile devices until their transmitted time intervals arrive. The link delay on the wireless links is assumed to be zero. Thus, we have
\begin{equation}
\Phi_{v_0}^{v_1}(a) = a,\quad
\phi_{v_0}^{v_1}(a) = 0, \quad \forall a\in\mathcal{N}_{\rm{tti}}
\end{equation}
where $v_0 \in\mathcal{U}$ and $v_1\in\mathcal{V}_{\rm{ap}}$ are the mobile users and APs on the s-path $p$.

\begin{figure}[ht]
\centering
\includegraphics[width=.80\linewidth]{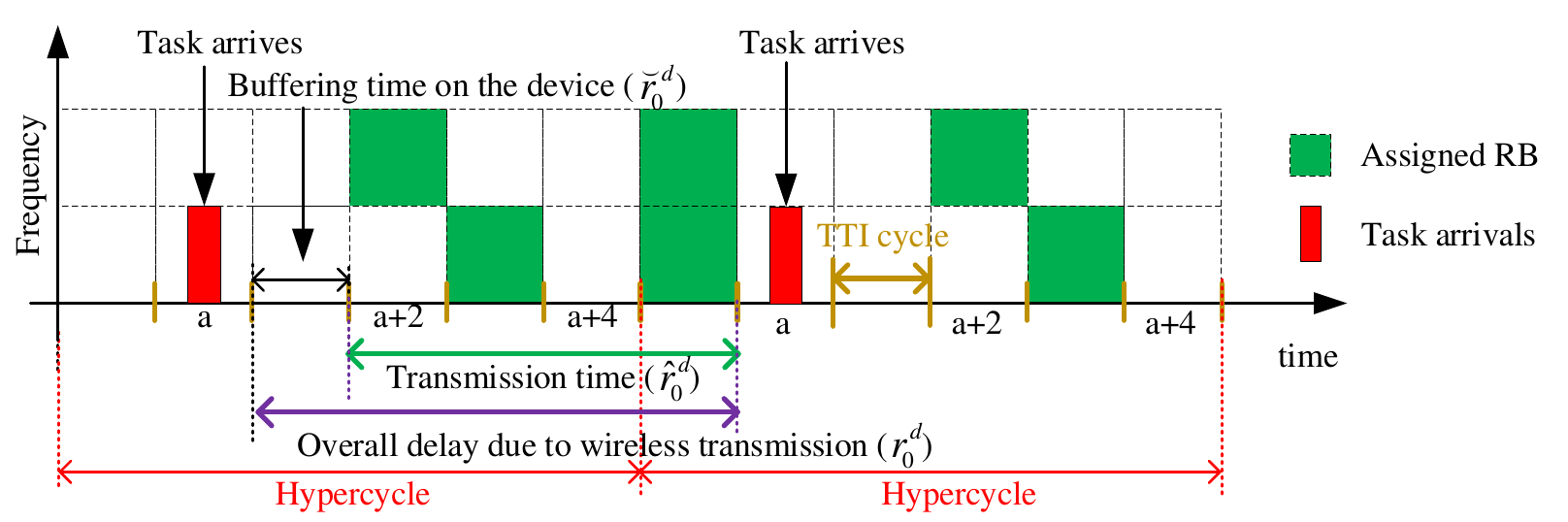}
\caption{\small Shift $r_0^d$ describes the overall delay due to the wireless transmission.}
\label{fig:time-lines}
\end{figure}

If the assignment ${\bf{y}}^d$ is decided, we use $r_0^d = {\breve{r}}_0^d + {\hat{r}}_0^d$ to denote the overall delay of the wireless transmissions. ${\breve{r}}_0^d$ (expressed in TTI cycles) is the buffering time on the mobile device and ${\hat{r}}_0^d$ (expressed in TTI cycles) is the wireless transmission time.

{\textit {Shifting and mapping at the edge of the RAN}}: The APs are equipped with both the TTI clock and DIP clock.
Since $\Delta_{\rm{tti}} > \Delta_{\rm{dip}}$ in most cases, we carry out the cycle shifting in terms of $\Delta_{\rm{dip}}$ for a better resource utilization.
An additional delay $r^d_1 \Delta_{\rm{dip}} + \phi_{v_1^-}^{v_1^+}(a)$ is introduced, where $\phi_{v_1^-}^{v_1^+}(\cdot)$ is the cycle mapping delay from the TTI cycles to the DIP cycles on AP $v_1$, and $a \in\mathcal{N}_{\rm{tti}}$ is the TTI index of last occupied RBs. The cycle mapping from the AP to its neighboring routers is considered taken placed in the same time domain (WN-D domain) since the DIP clocks have synchronized frequency.

\subsection{An end-to-end example}

\begin{figure}[ht]
\centering
\includegraphics[width=.99\linewidth]{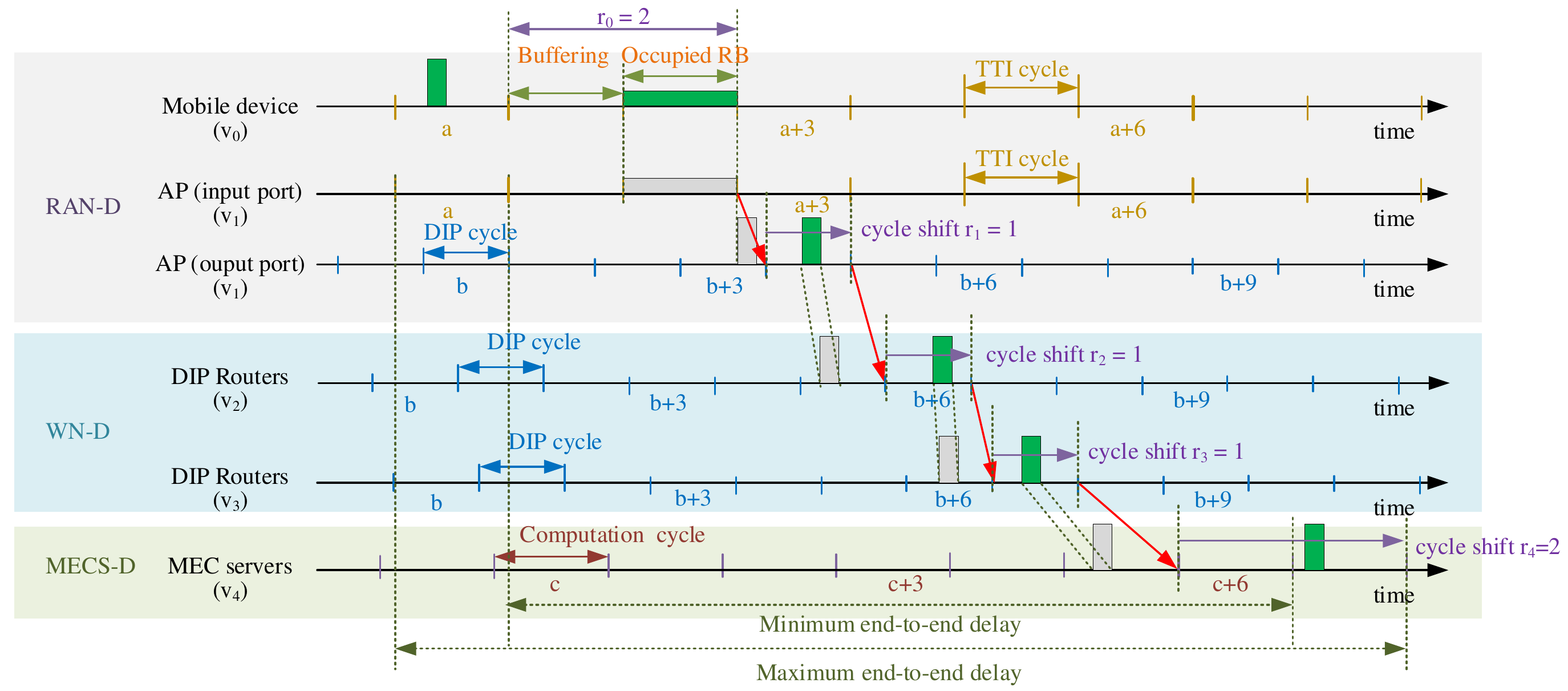}
\caption{\small An end-to-end example in the D-MEC networks. Demand $d$ uses the s-path: $p^d = \{v_0, v_1, v_2, v_3, v_4\}$. The jitter is produced at the first hop and the last hop. }
\label{fig:time-lines}
\end{figure}

Here, we depict an example. As Fig.~\ref{fig:time-lines} illustrated, the five-tuple of demand $d$ is described by $\left<s^d = v_0, T^d = \Delta_{\rm{hc}}, c^d = a, \omega^d, \Gamma^d\right>$, i.e., the application on mobile device $v_0$ generates a task (with total size $\omega^d$) at TTI cycle $a$ for every realization of the hypercycle. Demand $d$ is associated with the s-path $p^d = (v_0, v_1, v_2, v_3, v_4)$.

Assigned with the RBs at cycle $a+2$, the packets of demand $d$ would be temporally buffered in local for one TTI cycle (i.e., ${\breve{r}}_0^d=1$). With ${\hat r}^d_{0} = 1$ and $\Phi_{v_0}^{v_1}(a+2) = a+2$, AP $v_1$ finishes the receiving by the end of TTI cycle $a+2$. Then, given $\Phi_{v_1^-}^{v_1^+}(a+2)=b+3$ and $r_1^d = 1$, AP would re-send out the task at DIP cycle $b+4$. With mapping $\Phi_{v_1}^{v_2}(b+4) = b+5$, Router $v_2$ finishes the receiving by the end of (DIP) cycle $b+5$. The routers in WN-D adopt the DIP technique. Thereby, the cycle shift on Router $v_2$ and $v_3$ is always 1. Given the s-path and the shift values, MEC servers would finally receive the task before the end of computation cycle $c+5$. Given the shift $r_4^d = 2$, the server would finish the task before the end of cycle $c+7$.

{\textit {Service latency analysis:}} The maximum jitter on service latency is independent of the communication distance $|p^d|$, and always be a constant ($\Delta_{\rm{tti}} + \Delta_{\rm{mec}}$). The uncertainty brought by (i) heterogeneous cycle length, and (ii) offset on the start time of cycles is absorbed by the proposed cycle mapping. Given the cycle mapping and cycle shifting along the s-path, the worst-case transmission delay on each hop is determined, hence providing a delay-bounded MEC service (i.e., deterministic service).

\section{Problem formulation}

\subsection{Decision-making variables}
The following decision-making variables need to be jointly considered to achieve the deterministic MEC services.

{\textit {Admission control:}} We use $x^d$ to describe whether demand $d$ is accepted ($x^d=1$) or rejected ($x^d=0$). Let $\mathcal{X} = \{x^d\}_{d\in\mathcal{D}}$.

{\textit {MEC server selection and path determination:}} A s-path $p^d = (v_0,\cdots,v_{|p|}) \in \mathcal{P}$ is decided for every accepted demand $d$, where $(v_1, \dots, v_{|p|})$ is the transmission path and $v_{|p|}$ is the selected MEC server. Let $\mathcal{S} = \{p^d\}_{d\in\mathcal{D}}$.

{\textit {RB assignment:}} The RB assignment of demand $d$ is given by ${\bf y}^d = ({y}^d_{c,f})_{c\in\mathcal{N}_{\rm{tti}}, f\in\mathcal{F}}$, which also determines the values of $\{r_0^d\}_{d\in\mathcal{D}}$.

{\textit {Tasks forwarding control (cycle shifting):}} Along with the s-path $p^d$, we define an integer sequence ${\bf r}_p^d = (r_{0}^d, \cdots,  r_{|p^d|}^d)$, where $r_i^d$ is the cycle shift on the intermediate node $v_i$. We have $ r^d_i = 1, \forall i\in\{2, \dots, |p^d|-1\}$. Let $\mathcal{R} = \{{\bf r}^d_p\}_{d\in\mathcal{D}}$.

\subsection{Constraint on service latency}

The service latency of demand $d$, denoted as $\Delta^d(p^d, {\bf{r}}^d_p)$. If demand $d$ is accepted, i.e., $x^d=1$, then $\Delta^d(p^d, {\bf{r}}^d_p) \leq \Gamma^d$.

Consider an accepted demand $d$ with its associated s-path $p^d = \{v_0, \cdots, v_{|p^d|}\}$ and a shifting vector ${\bf r}_{p}^d$. In every realization of hypercycle, the intermediate node $v_i$ on the s-path $p^d$ re-send the demand $d$'s tasks out at cycle $c_i^d$, where $c_i^d$ is given by
\begin{equation}\label{e1}
c_i^d = \left\{ {\begin{array}{*{20}{l}}
{\rm{mod}}\left( { {\Phi _{v_1^ + }^{v_1^ - }\left( {c_0^d} \right) + r_1^d}, {N_{{\rm{dip}}}}} \right), &{i = 1}\\
{\rm{mod}}\left( { {\Phi _{{v_{i - 1}}}^{{v_i}}\left( {c_{i - 1}^d} \right) + r_i^d}, {N_{{\rm{dip}}}}}\right), &{i \in \left\{ {2, \cdots ,|p{|^d} - 1} \right\}}\\
{\rm{mod}}\left( { {\Phi _{{v_{i - 1}}}^{{v_i}}\left( {c_{i - 1}^d} \right) + r_i^d}, {N_{{\rm{mec}}}}}\right), &{i = |p{|^d}}
\end{array}} \right.
\end{equation}
where $c^d_0 = {\rm{mod}}\left({c^d} + r_0^d,  N_{{\rm{tti}}}\right)$ and $\Phi _{v_1^ + }^{v_1^ - }\left( {c_0^d} \right)$ is the cycle mapping function from TTI cycles to DIP cycles on AP $v_1$.

We denote as $\Delta_{i}^d$ the (maximum) accumulated delay for demand $d$ to be transmitted on intermediate node $v_i$. It is easily calculated as
\begin{equation}\label{e1}
{\Delta _i ^d} = \left\{ {\begin{array}{*{20}{l}}
{(1+r_d^0){\Delta _{{\rm{tti}}}} + \phi _{v_1^ + }^{v_1^ - }(c_0^d) + r_d^1{\Delta _{{\rm{dip}}}}}&{i = 1}\\
{{\Delta _{i - 1}^d} + \phi _{{v_{i - 1}}}^{{v_i}}(c_{i - 1}^d) + r_i^d{\Delta _{{\rm{dip}}}}}&{i = \left\{ {2, \cdots ,|p{|^d} - 1} \right\}}\\
{{\Delta _{i - 1}^d} + \phi _{{v_{i - 1}}}^{{v_i}}(c_{i - 1}^d) + r_i^d{\Delta _{{\rm{mec}}}}}&{i = |p{|^d}}
\end{array}} \right.
\end{equation}
where $\phi _{v_1^ + }^{v_1^ - }(c_0^d)$ is the mapping delay on AP $v_1$. The constraints on the service latency for demand $d$ is then given by
\begin{equation}\label{c1}
\Delta^d (p^d, {\bf{r}}^d_p) \leq \Delta_{|p^d|} ^d \leq \Gamma^{d}, \quad \forall d\in\mathcal{D}.
\end{equation}

\subsection{Constraint on resource capacities}
Dedicated resources are reserved for the exclusive use of the tasks. The traffic aggregated on the resources could not exceed their maximum capacity (i.e., RB capacity, link bandwidth, CPU cycles).

{\textit{Constraints on the RB capacity:}} The total capacity of assigned RBs for demand $d$ should exceed its amount of traffic, represented as
\begin{equation}\label{c2}
\begin{aligned}
\sum\limits_{c = c^d + \breve{r}_0^d}^{c^d + {r}_0^d} \sum\limits_{f\in\mathcal{F}} y _{c, f}^{d} {\rm{BW}}_{c, f}^{\rm{res}} \geq \omega^d, \quad \forall d \in \mathcal{D}
\end{aligned}
\end{equation}
Meanwhile, we have the condition that each RB can only be assigned to a single demand at a time, i.e.,
\begin{equation}\label{c3}
\begin{aligned}
\sum\limits_{d\in \mathcal{D}} y _{c, f}^{d} \leq 1,\quad \forall c \in\mathcal{N}_{\rm{tti}}, f\in\mathcal{F}
\end{aligned}
\end{equation}

{\textit{Constraints on the link bandwidth}: }
For every realizations of the hypercycle, demand $d$ always consumes a certain capacity $w_{e}^d(c)$ on wired link $e=(v_{i-1}, v_{i})$ along its s-path $p^d$. Here, $c$ is the index of DIP cycle, and $i\in\{2,\cdots,|p^d|-1\}$. $w_{e}^d(c)$ is given by
\begin{equation}
w_{e}^d\left( c \right) = \left\{ {\begin{array}{*{20}{l}}
{{\omega ^d}}&{if\;\;c = c_{i - 1}^d}\\
0&{otherwise}
\end{array}} \right.
\end{equation}
Then, the aggregate traffic at cycle $c$ on link $e=(v_{i-1}, v_{i})$ is given by $
w_{e}(c) = \sum\nolimits_{d\in\mathcal{D}} x^d w_{e}^d\left( c \right)$.
The constraints on the link bandwidth can be given by
\begin{equation}\label{c4}
\begin{aligned}
w_{e}(c) \leq  {\rm{BW}}^{\rm{link}}_e \cdot \Delta_{\rm{dip}}, \quad \forall e \in \mathcal{E}_{\rm{wired}}, c \in \mathcal{N}_{\rm{dip}}
\end{aligned}
\end{equation}

{\textit{Constraints on the computational resources}: } Demand $d$ also consumes a certain resources on its associated MEC server $v_{|p^d|}$ (indicated by its s-path $p^d$) for every realization of the hypercycle, given by
\begin{equation}
w_{v_{|p^d|}}^d\left( c \right) = \left\{ {\begin{array}{*{20}{l}}
{{\omega ^d} \kappa}&{if\;\;c = c_{|p^d|}^d}\\
0&{otherwise}
\end{array}} \right.
\end{equation}
Then, the aggregate traffic at cycle $c$ on server $v_{|p^d|}$ is given by $
w_{v_{|p^d|}}(c) = \sum\nolimits_{d\in\mathcal{D}} x^d w_{v_{|p^d|}}^d\left( c \right)$. The constraints on the resource consumption of server $v_{|p^d|}$ can be given by
\begin{equation}\label{c5}
\begin{aligned}
w_{v_{|p^d|}}(c) \leq  {\rm{BW}}^{\rm{mec}}_{v_{|p^d|}} \cdot \Delta_{\rm{mec}}, \forall v_{|p^d|} \in \mathcal{V}_{\rm{mec}}, c \in \mathcal{N}_{\rm{mec}}
\end{aligned}
\end{equation}

\subsection{Objective function}

The aim of the controller is to accept a subset of demands $\mathcal{D}$ such that the total number of the accepted demands is maximized. Then, the problem of interest is given by:

\begin{subequations}\label{P1}
\begin{alignat}{2}
\mathcal{W}_{\rm{opt}} = \max\limits_{{\mathcal X}, {\mathcal{Y}}, {\mathcal{S}}, {\mathcal{R}}} & \sum_{d\in\mathcal{D}} x^d\\
\mbox{s.t.} \quad  & x^d \in \{0, 1\}, \quad \forall d\in\mathcal{D} \label{c6}\\
 & |p^d| \leq H, \quad \forall d\in\mathcal{D} \label{c7} \\
 & \eqref{c1},\eqref{c2},\eqref{c3},\eqref{c4},\eqref{c5} \notag
\end{alignat}
\end{subequations}
Constraint \eqref{c6} gives the self-explanatory of the admission control. Constraint \eqref{c7} ensures that the number of hops along the s-path is no greater than $H$ per path.

{\textit {NP-hardness}:}  Problem~\eqref{P1} is an NP-complete problem. The controller need to decides if, for a given threshold $l \in  \mathbb{R}_+$, there is a feasible solution to Problem~\eqref{P1} with objective value $\mathcal{W}_{\rm{opt}}\leq l$. The following reduction proof is based on the well-known $k$-Disjoint Paths (k-DP) problem~\cite{korte2009combinatorial}. We consider the (NP-complete) version of $k$-DP which decides if $k$ arc-disjoint paths can be found between nodes $s$ and $t$ in a directed graph $\mathcal{G}$. This problem can be reduced to an instance of Problem~\eqref{P1} by setting $N_{\rm{tti}} = N_{\rm{dip}} = N_{\rm{mec}} = 1$, $\omega^d = 1, \forall  d\in\mathcal{D}$, and $|\mathcal{V}_{\rm{mec}}|=1$ that all have source (i.e., mobile devices) and destination (i.e., the MEC server). The capacity of every link $e$ is chosen to be one. Choosing $k = l$, Problem~\eqref{P1} returns true if and only if there are $k$ arc-disjoint paths in the network $\mathcal{G}$. Since all reduction steps are polynomial in the problem size, the NP-hardness proof is complete. Furthermore, it is clear that  Problem~\eqref{P1} belongs to NP since the validity of any solution can be checked in polynomial time. Thus, Problem~\eqref{P1} is NP-complete.

\section{Simulation}
We evaluate the proposed D-MEC network with real-world networks produced from SNDlib~\cite{OrlowskiPioroTomaszewskiWessaely2010}. Problem~\eqref{P1} is solved with a Tabu search-based algorithm~\cite{gendreau2005tabu}. The network is built in the OMNet++, and consists of 10 aggregate routers and 5 core routers. By default, we place 10 APs that access the core routers through the aggregate routers. 5 MEC servers (including 3 edge servers and 2 central servers) are installed, linked to aggregate routers and core routers, respectively. The link delays among the routers vary from 30~$\mu$s to 60~$\mu$s, randomly.
The capacities of links are set as 10 Gbps, uniformly. At the APs and MEC servers, we install 20 cyclic queues. In the RAN, the mobile devices are dispersed randomly around their associated APs with the communication range shorter than 20 meters. We consider
the Gaussian channel model for the wireless links with white noise equalling -120dBm. The RAN operates with TTI duration $\Delta_{\rm{tti}} = 125$ $\mu$s.

In the simulations, each task is encapsulated as a series of packets with total size $\omega^d = 1$~KB. The duration of demands $T^d$ is set as 1 ms. We set the cycle durations in different domains as $\Delta_{\rm{dip}} = 15$ $\mu$s, $\Delta_{\rm{mec}} = 30$ $\mu$s, and $\Delta_{\rm{hc}} = 1$ ms, respectively. We emulate the background flows in the network with the UDP burst applications in the OMNet++.

\begin{figure}[ht]
\centering
\includegraphics[width=.80\linewidth]{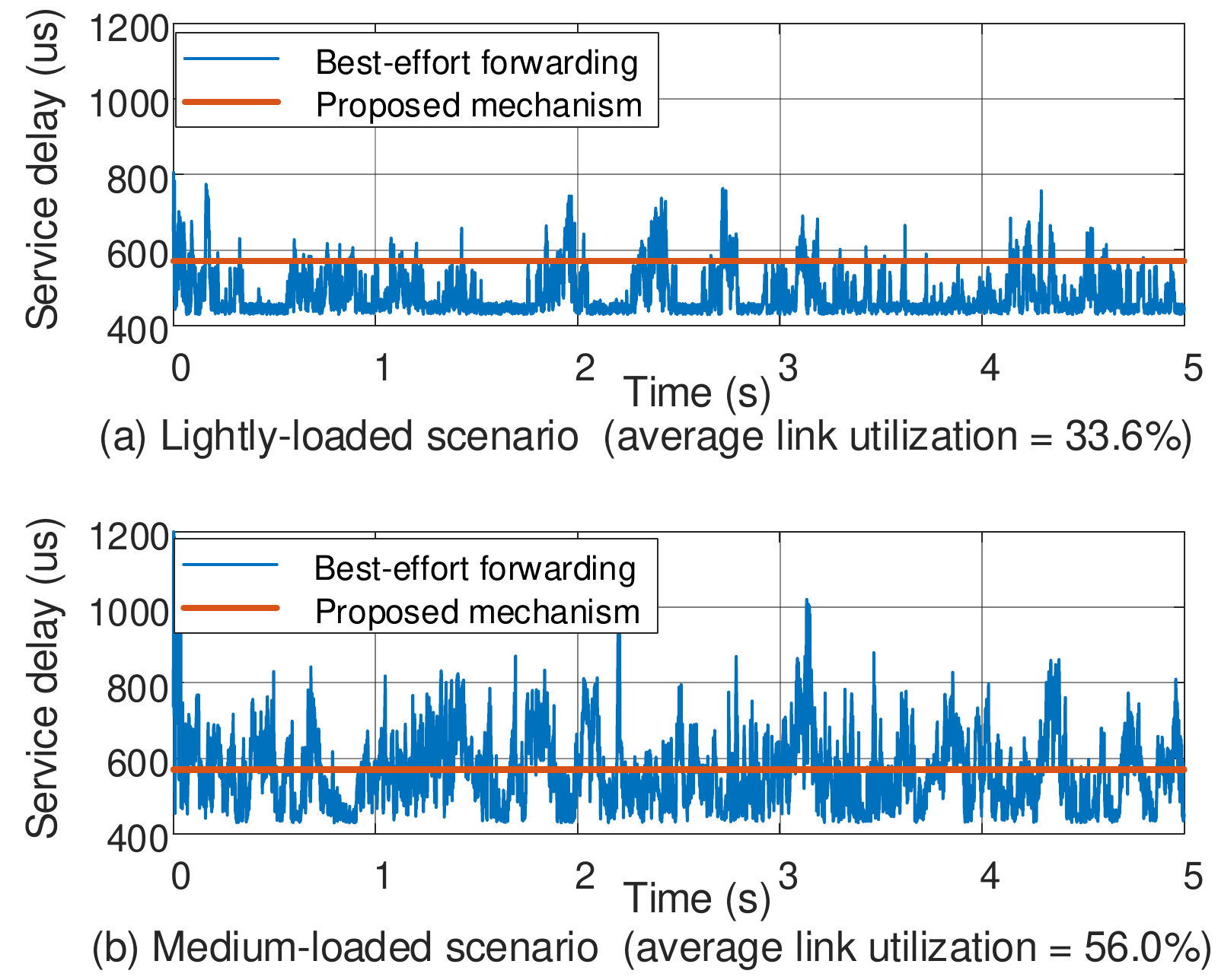}
\caption{\small Service latency with different level of congestions: (a) lighted-loaded scenarios, (b) medium-loaded scenarios.}
\label{fig:sim1}
\end{figure}
Fig.~\ref{fig:sim1} uses different levels of the UDP burst (reflected in the average link utilization) to evaluate the service latency in various congestion levels. The curves in red verify that the D-MEC network can achieve a deterministic service latency (around 570 $\mu$s). Meanwhile, significant fluctuations are experienced with the best-effort forwarding. The service latency varies from 440~$\mu$s (440~$\mu$s) to 800~$\mu$s (1.2~ms) in the lightly-loaded (highly-loaded) scenarios). The reason is that, with best-effort forwarding, multiple flows may be received in a small time period so that the burst data rate is tens or hundreds of times higher than the port bandwidth, i.e., traffic jams. On the other hand, the proposed D-MEC network regulates the traffic in each hop (with the cyclic-forwarding manner), hence eliminating the bursts and bringing the deterministic service latency. Nevertheless, the traffic regulation requires additional delays (owing to cycle mapping and cycle shifting). The minimum service latency in the D-MEC is larger than the cases in the best-effort forwarding.

\begin{figure}[ht]
\centering
\includegraphics[width=.80\linewidth]{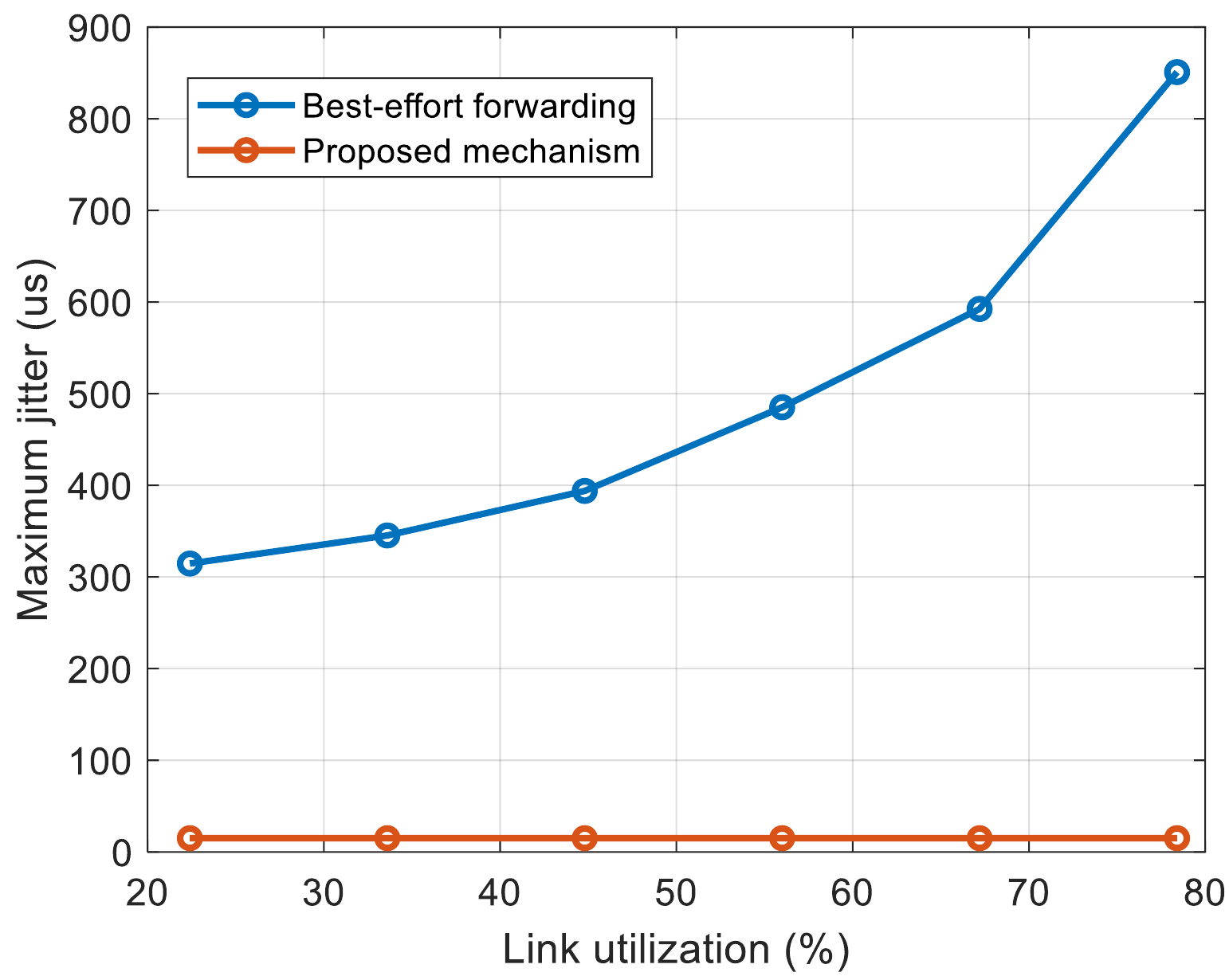}
\caption{\small  Maximum jitter with different level of congestions (measured by the average link utilization). }
\label{fig:sim2}
\end{figure}

Fig.~\ref{fig:sim2} depicts the performance on the maximum jitter. Here, we gradually increase the number of time-sensitive demands to produce different congestion levels. A deterministic service latency can be obtained with the proposed D-MEC network. Given the dedicated resources,  most jitters come from the resource competition among the demands that are assigned to the same time cycles (e.g., DIP cycles). Moreover, the cyclic mapping and shifting can bound the transmission latency per hop. On the other hand, with the best-effort forwarding, the demands not only compete with the other time-sensitive traffic, but also with the background flows, which leads to significant delay variations. Without time-based traffic regulation, ``microburst'' may occur.

\begin{figure}[ht]
\centering
\includegraphics[width=.80\linewidth]{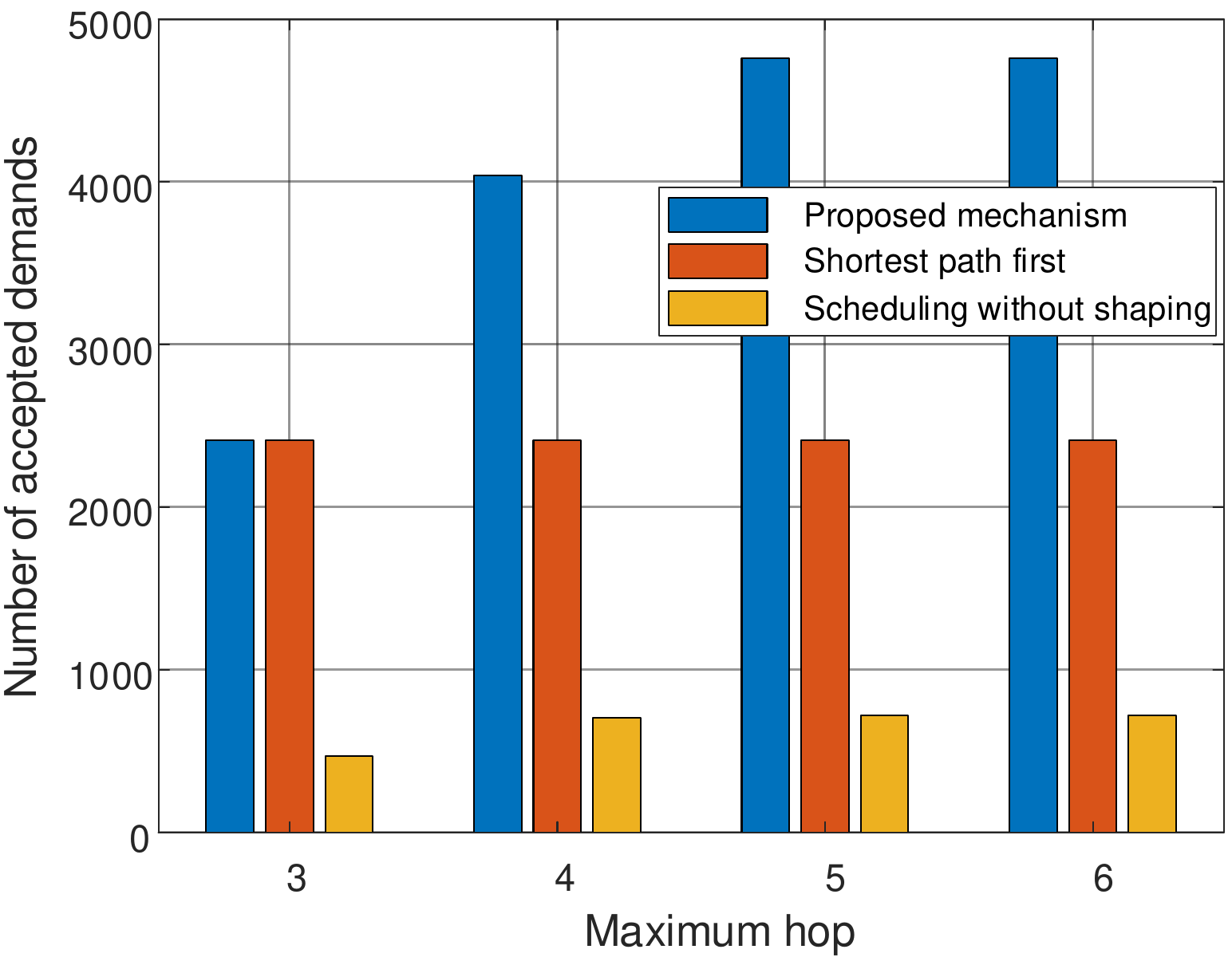}
\caption{\small $\mathcal{W}_{\rm{opt}}$ in problem~\eqref{P1} with different $H$ ($\Gamma^d = 1$~ms). Two baseline algorithms are used, i.e., ``shortest path first'' and ``scheduling without shaping''. ``Shortest path first'' uses the shortest path for transmission. ``Scheduling without shaping'' does not carry out the traffic shaping at the edges of the domains, i.e., $r^d_i = 1, \forall d, i$. }
\label{fig:sim3}
\end{figure}

Fig.~\ref{fig:sim3} depicts the advantages of the edge shaping (on the APs and MEC servers) and path selection with different maximum hops. The traffic shaping on the APs and MEC servers significantly improves the network throughput by comparing the results of the ``proposed mechanism'' and ``scheduling without shaping''. Given the heterogeneous underlaid resources, the proposed mechanism can balance the workload on the downstream links, hence improving resource utilization. Without shaping, traffic only aggregates on a few of the cycles pointed out by the cycle mapping mechanism. Meanwhile, the selection of transmission paths also contributes to throughput enhancement. With a larger $H$ (maximum hop), the traffic can be balanced on a bigger area. On the other hand, the greedy path selection in the ``shortest path first'' sometimes results in local optima.

\section{Conclusion}

This paper delves into the adoption of the DIP for the MEC networks to achieve the deterministic MEC service. We divide the network into different domains, i.e., RAN-D, WN-D, and MECS-D, for the heterogeneous resources (e.g., wireless RBs, link bandwidth, CPU cycles). A cycle mapping is used to enable seamless and deterministic transmissions across the domains with different settings. Moreover, we propose a cycle shifting mechanism, which shapes the traffic at the edges of domains (i.e., at the APs and MEC servers) to balance the workload on the downstream nodes and links, hence enhancing the network throughput. Extensive simulations corroborate the effectiveness of the proposed D-MEC network under various settings. With real-world network topologies generated by SNDlib, the proposed D-MEC network is shown to significantly outperform the traditional mechanisms (i.e., best-effort forwarding).



%


\end{document}